\documentclass[a4paper,fleqn,usenatbib]{mnras}

\usepackage{newtxtext,newtxmath}
\usepackage[T1]{fontenc}
\usepackage{ae,aecompl}

\usepackage[dvips]{graphicx}	% Including figure files
\usepackage{amsmath}	% Advanced maths commands
\usepackage{amssymb}	% Extra maths symbols

\usepackage{natbib}
\title[A fast method to identify mean motion resonances]{A fast method to identify mean motion resonances}

\author[E. Forg\'acs-Dajka et al.]{
E. Forg\'acs-Dajka$^{1,2}$,\thanks{E-mail: E.Forgacs-Dajka@astro.elte.hu}
Zs. S\'andor$^{1,3}$,
and B. \'Erdi$^1$
\\
$^1$Department of Astronomy, E\"otv\"os Lor\'and University, P\'azm\'any P\'eter s\'et\'any 1/A, H-1117 Budapest, Hungary\\
$^2$Wigner RCP of the Hungarian Academy of Sciences, 29-33 Konkoly-Thege Mikl\'os Str, H-1121 Budapest, Hungary\\
$^3$Konkoly Observatory, Research Centre for Astronomy and Earth Sciences, Hungarian Academy of Sciences, \\
H-1121 Budapest, Konkoly Thege Mikl\'os \'ut 15-17., Hungary
}

\date{Accepted 2018 March 7. Received 2018 March 7; in original form 2018 January 3}

\pubyear{2017}

\begin{document}
\label{firstpage}
\pagerange{\pageref{firstpage}--\pageref{lastpage}}
\maketitle

\begin{abstract}
The identification of mean motion resonances in exoplanetary systems or in the Solar System might be cumbersome when several planets and large number of smaller bodies are to be considered. Based on the geometrical meaning of the resonance variable, an efficient method is introduced and described here, by which mean motion resonances can be easily find without any a priori knowledge on them. The efficiency of this method is clearly demonstrated by using known exoplanets engaged in mean motion resonances, and also some members of different families of asteroids and Kuiper-belt objects being in mean motion resonances with Jupiter and Neptune respectively.
\end{abstract}

\begin{keywords}
celestial mechanics -- methods: numerical -- (stars:) planetary systems
\end{keywords}

%%%%%%%%%%%%%%%%%%%%%%%%%%%%%%%%%%%%%%%%%%%%%%%%%%

%%%%%%%%%%%%%%%%% BODY OF PAPER %%%%%%%%%%%%%%%%%%

\section{Introduction}
As a result of the ongoing discoveries, the number of exoplanets is continuously growing; up to now more than 3500 planets are known in more than 2600 planetary systems. These numbers indicate that a significant amount of planets form multi-planet systems, in which at least two planets are revolving around a central star. Planets form in gas rich protoplanetary disks, therefore beside gravity, forces arising from the ambient gaseous material play important role as altering typically their semi-major axes during a process called orbital migration. If the orbital distance between two planets decreases in time due to their migration, the planets can be orbitally locked at certain ratio of their semi-major axes, where the ratio of the mean angular velocities (mean motions) can be expressed as a ratio of small positive integer numbers. If certain dynamical conditions are fulfilled, a resonant capture may happen meaning that the ratio of the semi-major axes and that of the mean motions does not change during the further migration of the planets. 

Before the Kepler-era, mean motion resonances (MMRs) outside the Solar System have been detected between giant planets. One of the most remarkable resonant systems discovered at that time is around the M dwarf star GJ 876 in which two planets are in 2:1 MMR \citep{2001ApJ...556..296M}. Later on three planets in Laplace-type resonance have been identified \citep{2010ApJ...719..890R}, being the first known resonant chain of exoplanets resembling to the Jupiter's Galilean satellites Io-Europa-Ganymede. Other examples for the 2:1 MMR are the system HD~128311, in which two massive planets are in resonance but with only one critical argument librating \citep{2005ApJ...632..638V}, and HD~73526, also with one librating critical argument \citep{2006ApJ...647..594T}. In the system HD 60532 two giant planets are in 3:1 MMR \citep{2008A&A...491..883D}. A further example for a planetary system that might be in the Laplace resonance is the system HR 8799 in which three massive planets are orbiting around an A-type star \citep{2014MNRAS.440.3140G}. More recently, more planets have been discovered to be dynamically locked in chains of MMRs among the Kepler systems. In Kepler-60 three planets with masses around  $\sim 4M_\oplus$ are in a 5:4:3 Laplace-type MMR \citep{2016MNRAS.455L.104G}. The four Neptune-mass planets in Kepler-223 have periods in ratios close to 3:4:6:8 \citep{2016Natur.533..509M} being another clear indication of planetary migration.

Beside the exoplanetary systems, mean motion resonances also play an important role in shaping the dynamics of the Solar System bodies. Mean motion resonances in the Solar System usually occur between a planet and small bodies: e.g. the members of the Hilda group of asteroids are in a 3:2, while the Trojan asteroids are in a 1:1 MMR with Jupiter. MMRs between terrestrial planets and particular members of the asteroid family Hungaria can also be found. Moreover, as the existence of the Kirkwood gaps clearly indicates, the dynamical structure of the main belt is shaped by the MMRs between the asteroids and Jupiter (though their depletion is not only due to MMRs). Regarding larger objects, between Neptune and the dwarf planet Pluto there is a 3:2 MMR, which is a protective resonance keeping them on non-near-approaching orbital positions. This is of high importance, since due to the high eccentricity of Pluto's orbit ($e = 0.25$) the projection of its perihelion to Neptune's orbital plane lies inside the orbit of Neptune thus the two orbits are crossing each other (in projection). Interestingly, there are many other Kuiper-belt objects that are locked in the 3:2 MMR with Neptune, these are the plutinos sharing their orbits with Pluto.

When studying the dynamics of exoplanetary systems with planets engaged in chains of mean motion resonances, or the motion of small celestial bodies, asteroids or trans-Neptunian objects (TNOs), one often finds the problem how to identify the possible mean motion resonances among the bodies involved. For example, during the formation of the dynamical structure of the trans-Neptunian region, resonances swept through the primordial disk due to the migration of the planets, and planetesimals were captured into mean motion resonances. Modelling this process, a monitoring is necessary to see how the captures take place, which planetesimals are captured to which resonances \citep{2005AJ....130.2392H,2008Icar..196..258L}. Similarly, studying the dynamical evolution of an asteroid family, one has to check possible resonances with the planets \citep{2013P&SS...84....5G}. This needs the computation of the resonance variables, to check their possible libration, for all reasonable (often very high order) resonant ratios of the mean motions of all small bodies with all considered planets, and this has to be repeated several times during the investigated time-span resulting in a considerable computational workload. 

In order to make easier the identification of the different mean motion resonances, we present a method we name \emph{FAIR} (as \emph{FA}st \emph{I}dentification of mean motion \emph{R}esonances) that is easy to use and by which the identification of the MMRs is possible without any a priori knowledge on them. Our paper is organized as the following: in the next section we introduce \emph{FAIR} for inner and outer type MMRs, then in two further sections particular examples are presented for Solar System objects, and for known exoplanetary systems, also including the study of resonant chains of planets.    

\section{The method \emph{FAIR}}

Let us denote by $a$, $a'$ the semi-major axes of two celestial bodies locked in a MMR, and by $n$, $n'$ the corresponding mean motions. For the sake of simplicity let us deal with the situation of an asteroid and a planet, in this case the primed quantities refer to the planet. This notation enables us to distinguish between inner and outer resonances from the point of view of the asteroid. In the case of two giant planets neither of the bodies are in such distinguished positions, the non-primed quantities refer to the planet being the body to which respect the MMR is studied.

Considering an asteroid and a planet, they can be either in
\begin{description}
\item {(i) \emph{inner} mean motion resonance, 
if $a<a'$ and the ratio of the mean motions is
approximately
\begin{equation}
       \frac{n}{n'}=\frac{p+q}{p},
\end{equation} or in}
\item{(ii) \emph{outer} mean motion resonance, if $a>a'$ and approximately
\begin{equation}
       \frac{n'}{n}=\frac{p+q}{p},
\end{equation}  }     
\end{description}
where $p$ and $q$ are relative prime integers. Here $q$ is the order of the resonance. The resonant perturbations are proportional to the $q$th power of the orbital eccentricities, thus for small eccentricities  low order resonances are the most important. However, for large eccentricities high order resonances can also be significant. (These refer to the so-called eccentricity-type resonance, when the longitudes of the perihelion are also involved, as in the cases studied below.)

Resonances can be studied by using resonance variables. For a given mean motion resonance there can be several types of resonance variables \citep{1999ssd..book.....M}. From among them we consider those for which the method \emph{FAIR} is applicable.

\subsection{Inner resonance}
Since beside the selected asteroid and planet there are also other planets in a planetary system, we should also take into consideration the secular rates $\dot{\varpi}$, $\dot{\varpi}'$ of the perihelion longitudes, caused by the mutual gravitational perturbations. 

In this case one can obtain the following resonance variables:
\begin{equation}
  \theta_1 \equiv (p+q) \lambda' - p \lambda -q \varpi,
  \label{eq6}
\end{equation}
\begin{equation}
  \theta_2 \equiv (p+q) \lambda' - p \lambda -q \varpi',
  \label{eq7}
\end{equation}
where $\lambda=M+\varpi$, $\varpi=\omega+\Omega$, $\lambda$ is the mean orbital longitude, $M$ the mean anomaly, $\omega$ the argument of perihelion, $\Omega$ the longitude of the ascending node, $\varpi$ the longitude of the perihelion of the asteroid. The primed variables refer to the planet with similar meaning. 

Eqs (\ref{eq6}) and (\ref{eq7}) refer to the case of exact resonance. Near a resonance, Eqs (\ref{eq6}) and (\ref{eq7}) are satisfied approximately, and the resonance variables oscillate (librate) around a mean value, which is usually  $0^{\circ}$ or $180^{\circ}$, with amplitudes $\Delta \theta < 180^{\circ}$.

Actually, in the neighbourhood of a resonance, the behaviour of a resonance variable can be quite complex. Depending on the actual system and initial conditions, it can exhibit libration, circulation, alternation between libration and circulation, or chaotic fluctuations. When searching for resonance, our aim is to establish that the resonant ratio of the mean motions (or their combinations with the secular frequencies) is maintained and the corresponding resonance variable librates. These are necessary and sufficient conditions that a resonance exist.  We note that the above introduced resonance variables appear as critical arguments in the series expansion of the perturbed two-body potential in terms that contain the $q\mathrm{th}$ powers of $e$ and $e^{\prime}$. There are mixed types of eccentricity-type resonances too, corresponding to critical arguments (and resonance variables) that contains coefficients of $ee^{\prime}$ type. In this work, however, we do not consider the latter cases of mixed eccentricity-type resonances.

Now let us consider Eq. (\ref{eq6}) in two cases. First, when $\lambda=\lambda^{\prime}$, that is at the conjunction of the asteroid and the planet, it follows from (\ref{eq6}) that
\begin{equation}
  \theta_c=q(\lambda -\varpi)=qM,
\label{eq_4_conj}
\end{equation}  
where $\theta_c$ is the value of $\theta_1$ at conjunction. Writing $\bar{\theta}=\theta_{10}$ and denoting the amplitude of $\theta_1$ by $\Delta \theta$, since 
\begin{displaymath}
\bar{\theta}-\Delta \theta \leq   \theta_c  \leq  \bar{\theta}+\Delta \theta ,
\end{displaymath}
it follows that
\begin{displaymath}
\bar{\theta}-\Delta \theta \leq   qM \leq  \bar{\theta}+\Delta \theta.
\end{displaymath}
Here $M$ is the mean anomaly of the asteroid at the moment of conjunctions, and $qM$ should be taken mod $2\pi$. Thus
\begin{displaymath}
\frac{\bar{\theta}+k2\pi-\Delta \theta}{q} \leq   M \leq  \frac{\bar{\theta}+k2\pi+\Delta \theta}{q},
\end{displaymath}
where $k$ is an integer. This gives $q$ centres 
\begin{displaymath}
\frac{\bar{\theta}}{q}+k \frac{2\pi}{q}, \quad k=0,1,\ldots,q-1
\end{displaymath} 
along the orbit of the asteroid around which the conjunctions of the asteroid with the planet can take place within regions of half-size $\Delta \theta/q$. For example, for $q=3$ and 
$\bar{\theta}=0^{\circ}$ the centres are at $0^{\circ}$, $120^{\circ}$, and 
$240^{\circ}$, while for $\bar{\theta}=180^{\circ}$ these are at $60^{\circ}$, $180^{\circ}$, and $300^{\circ}$.

In the second case, when $M=0^{\circ}$, that is at the perihelion of the asteroid, $\varpi=\lambda$, and it follows from (\ref{eq6}) that 
\begin{displaymath}
\theta_p=(p+q)(\lambda'-\lambda),
\end{displaymath} 
where $\theta_p$ is the value of $\theta$ at the perihelion. Since
\begin{displaymath}
\bar{\theta}-\Delta \theta \leq   \theta_p  \leq  \bar{\theta}+\Delta \theta ,
\end{displaymath}
it follows that
\begin{displaymath}
\bar{\theta}-\Delta \theta \leq (p+q)(\lambda'-\lambda)   \leq  \bar{\theta}+\Delta \theta.
\end{displaymath}
As before, it also follows that
\begin{displaymath}
\frac{\bar{\theta}+k 2\pi-\Delta \theta}{p+q} \leq  \lambda' - \lambda \leq  \frac{\bar{\theta}+k 2\pi+\Delta \theta}{p+q}
\end{displaymath}
giving $p+q$ centres 
\begin{displaymath}
\frac{\bar{\theta}}{p+q}+k \frac{2\pi}{p+q}, \quad k=0,1,\ldots,p+q-1
\end{displaymath} 
around which in regions with half-size $\Delta \theta/(p+q)$ the asteroid can be found with respect to the planet when the former is at perihelion.

The previous considerations can be summed up in stating that plotting $\lambda'-\lambda$ against $M$ in a rectangular coordinate system with $M$ as horizontal and $\lambda'-\lambda$  as vertical axis, there will be $q$ centres on the horizontal, and $p+q$ centres on the vertical axis. 

The same reasoning can be repeated with Eq. (\ref{eq7}), concluding to that plotting $\lambda'-\lambda$ against $M'$ there will be $q$ centres on the horizontal, and $p$ centres on the vertical axis.

\subsection{Outer resonance}

In the case of an outer resonance, the resonance variables for the eccentricity-type resonances are
\begin{equation}
  \theta_3=(p+q) \lambda - p \lambda' -q \varpi,
  \label{eq11}
\end{equation}
and
\begin{equation}
  \theta_4=(p+q) \lambda - p \lambda' -q \varpi'.
  \label{eq12}
\end{equation}
By similar considerations as before, it follows from Eq. (\ref{eq11}) that plotting 
$\lambda-\lambda'$ against $M$, there will be $q$ centres on the horizontal, and $p$ centres on the vertical axis. In the case of Eq. (\ref{eq12}), plotting $\lambda-\lambda'$ against $M'$, there will be $q$ centres on the horizontal, and $p+q$ centres on the vertical axis. Table \ref{tab1} gives a summary of the different cases in the inner and outer resonances.

\subsection{Inclination-type resonances of inner and outer type}
We have investigated the eccentricity-type resonances so far, however, it can be easily shown that with minor modifications, the method \emph{FAIR} also works for inclination-type resonanances. Let us consider the non-mixed inclination-type resonances of order $q$, where the critical arguments appear in terms containing the $q$th power of either $I$ or $I'$ (being the inclinations of the body and of the perturber, respectively) as coefficients in the series expansion of the perturbing potential:
\begin{equation}
  \theta_{I,1} = (p+q) \lambda' - p \lambda -q \Omega,
  \label{eq_inn_inc1}
\end{equation}
\begin{equation}
  \theta_{I,2} = (p+q) \lambda' - p \lambda -q \Omega',
  \label{eq_inn_inc2}
\end{equation}
where $\Omega$ and $\Omega'$ are the longitude of nodes of the body and the perturber, respectively. Equating the mean longitudes, $\lambda=\lambda'$, similarly to Equation \eqref{eq_4_conj}, conjunctions happen when
\begin{equation}
\theta_{I,c} = q(\lambda-\Omega)=q(\omega+M).
\end{equation}
This suggests that in the inclination-type resonances one should replace $M$ with $M+\omega$, and by a similar argumentation to the eccentricity-type resonances, there will be $q$ centres on the horizontal, and $(p+q)$ centres on the vertical axis when plotting $\lambda'-\lambda$ versus $M+\omega$ in a rectangular coordinate system. The above considerations can also be applied to the outer inclination-type resonances, when $\lambda'-\lambda$ should be plotted either versus $M+\omega$ or $M'+\omega'$. Thus by replacing in Table \ref{tab1} $M$ and $M'$ with $M+\omega$ and $M'+\omega'$, moreover $\varpi$ and $\varpi'$ with $\Omega$ and $\Omega'$ respectively, one can easily obtain the resonance variable of the corresponding inclination-type resonance.

\begin{table}
\caption{Resonance variables. The third column shows the variables to be plotted versus each other, the fourth and fifth the number of centres on the horizontal and vertical axes.}
\label{tab1}
\begin{tabular}{ccccc}
\hline
type & resonance variable & plot & hor & vert \\
 \hline
 inner & $(p+q) \lambda' - p \lambda -q \varpi $ &  $\lambda'-\lambda$ vers $M$ & $q$ & $p+q$ \\
 inner & $(p+q) \lambda' - p \lambda -q \varpi' $ & $\lambda'-\lambda$ vers $M'$ & $q$ & $p$ \\
 outer & $(p+q) \lambda - p \lambda' -q \varpi $  &  $\lambda-\lambda'$ vers $M$ & $q$ & $p$ \\ 
 outer & $(p+q) \lambda - p \lambda' -q \varpi' $ &  $\lambda-\lambda'$ vers $M'$ & $q$ & $p+q$ \\ 
 \hline
\end{tabular}   
\end{table}

\section{Identification of resonances}

The properties of the resonance variables, described in Section 2, can be used to find easily whether a resonance exists between a planet and an asteroid. Deciding from the semi-major axes that the looked for resonance is inner or outer, then searching for inner resonances, one plots $\lambda'-\lambda$ versus $M$ and $M'$, (or in the cases of inclination-type MMRs versus $M+\omega$ and $M'+\omega'$) computed by numerical integration of the equations of motion for sufficiently long time. If there is a resonance, some stripes will appear on the plots (see the examples below),  and counting the number of their intersections with the horizontal and vertical axis will provide the $p$ and $q$ values of the resonance. Searching for outer resonances, the procedure is the same with plotting  $\lambda-\lambda'$ versus $M$ and $M'$ and counting the numbers of intersections on the horizontal and vertical axis. Since in the case of a resonance the appearance of the proper number of intersections is a necessary condition, once $p$ and $q$ are found, the ratio of the mean motions and the libration of the resonance variable should be checked with them.

When processing the above plots, there can be raised two technical questions: (i) how long the numerical integrations should be carried out, and (ii) how frequently should the points be plotted in order to obtain good quality plots suitable for determination of a MMR. Without giving a definite answer, we have empirically found that regarding the plotting frequency, the one hundredth of the period of the inner body certainly will give good result when using an integration time that corresponds to one hundred periods of the outer body.

\subsection{Solar System examples}

In this section we provide a few examples for the application of the method \emph{FAIR} to Solar System bodies involved in various mean motion resonances. The orbits have been numerically calculated by using an own-developed adaptive step size Runge-Kutta-Nystrom 6/7 N-body integrator \citep{1978CeMec..18..223D}. In our integrations all planets except Mercury have been taken into account, and  as initial conditions we used heliocentric coordinates and velocities taken from JPL Horizons at the epoch JD~2457754.50.

\subsubsection{153~Hilda - Jupiter 3:2 MMR}

\begin{figure}
\includegraphics[width=\columnwidth]{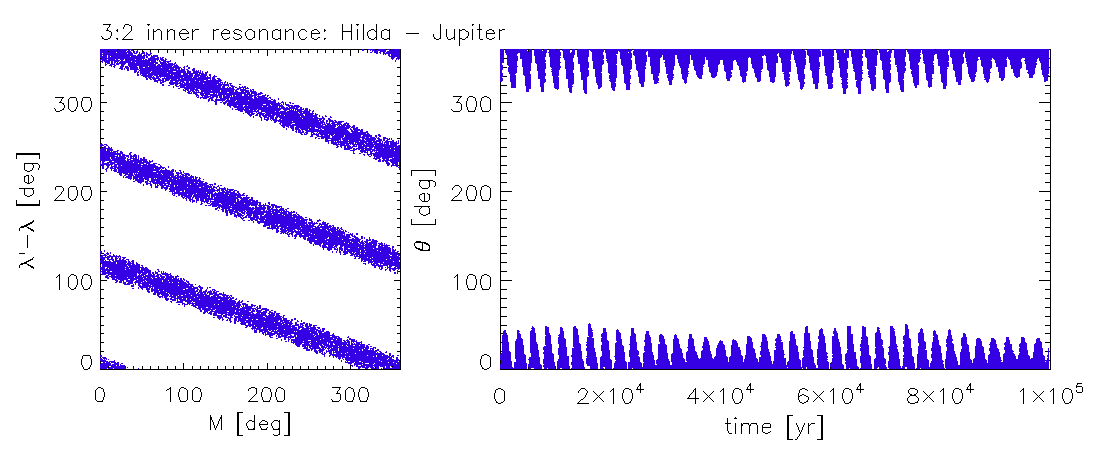}
\caption{Asteroid 153~Hilda in a 3:2 MMR with Jupiter.}
\label{fig:jupiterhilda}
\end{figure}

The main dynamical characteristic of the Hilda group of asteroids (named after 153~Hilda) is that its members are locked in a 3:2 MMR with Jupiter  \citep{2008MNRAS.390..715B}. As a first example on how the method \emph{FAIR} works is given by plotting $\lambda' - \lambda$ versus $M$ (see Figure \ref{fig:jupiterhilda}, left panel), where $\lambda'$  is the mean longitude of the perturbing body (Jupiter, in this case), and $\lambda$, $M$ are the mean longitude, and mean anomaly of the asteriod 153~Hilda, respectively. Since this is a MMR of inner type, and $\lambda' - \lambda$ is plotted against $M$, according to the first row of Table 1 the counting of the number of intersecting stripes with the horizontal and vertical axis gives $q = 1$ and $p+q=3$, see the left panel of Figure \ref{fig:jupiterhilda}. Thus, the ratio of the mean motions is $(p+q)/p = 3/2$, and the corresponding resonance variable (or resonant angle) is $\theta_1 = 3\lambda' - 2\lambda -\varpi$, librating in this case around $0^\circ$ as shown in the right panel of Figure \ref{fig:jupiterhilda}. We note that in this case, the ratio of the mean motions and the resonance variable are determined solely by counting the crossings of the stripes with the horizontal and vertical axes.

\subsubsection{Neptune - Pluto 3:2 MMR}

\begin{figure}
\includegraphics[width=\columnwidth]{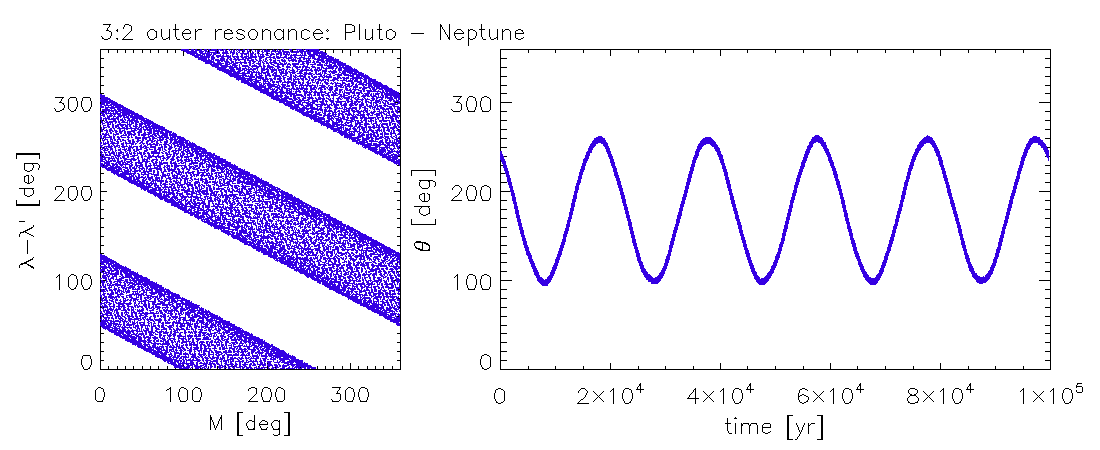}
\caption{The 3:2 MMR between Neptune and Pluto}
\label{fig:neptunepluto}
\end{figure}

As a further well known example, we consider the case of Neptune and Pluto, being in a 3:2 MMR \citep{1965AJ.....70...10C}. In this case, the more massive Neptune (the perturber) orbits closer to  the Sun, thus the MMR under study is of outer type. In the left panel of Figure \ref{fig:neptunepluto} we displayed $\lambda - \lambda'$ versus $M$. Since this is a MMR of outer type, according to the third row of Table 1 the counting of the crossings of the stripes with the horizontal and vertical axis gives $q=1$ and $p=2$. Thus the ratio of the mean motions is $(p+q)/p=3/2$, and the corresponding resonance variable, $\theta_3 = 3\lambda - 2\lambda' - \varpi$ librates around $180^\circ$. 

We note that although the above two well known mean motion resonances can be identified easily without the application of the method \emph{FAIR}, we were able to identify them without knowing their character. In what follows, we show some more sophisticated cases, in which the identification of the resonances and the corresponding resonance variables might be more complicated.   

\subsubsection{Neptune - 2005~TN$_\mathrm{74}$ 5:3 MMR}

\begin{figure}
\includegraphics[width=\columnwidth]{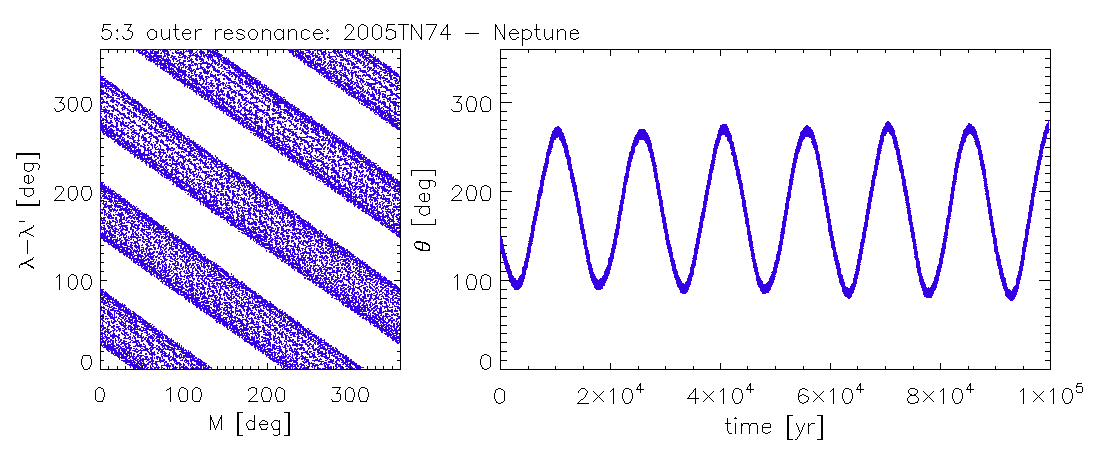}
\caption{The 2005TN$_\mathrm{74}$ trans-Neptunian object in a 5:3 MMR with Neptune.}
\label{fig:2005TNNeptune}
\end{figure}

2005~TN$_\mathrm{74}$ is a trans-Neptunian object (TNO) discovered by \cite{2005MPEC....U...97S}. In order to demonstrate the effectiveness of the method \emph{FAIR}, let us forget for now that this body is in a 5:3 MMR with Neptune. The only information we suppose to have is that this TNO orbits outside Neptune, therefore when seeking for a resonant behaviour we should consider a MMR of outer type. To do so we display the plot $\lambda-\lambda'$ versus $M$, where $\lambda'$ is the mean longitude of Neptune, while $\lambda$ and $M$ are the mean longitude and mean anomaly of 2005~TN$_\mathrm{74}$, respectively. Studying the left panel of Figure \ref{fig:2005TNNeptune}, we can count the numbers of crossings of the stripes with the axes, yielding $q=2$ and $p=3$ (see the third row of Table \ref{tab1}). Thus the ratio of the mean motions is $(p+q)/p=5/3$, and the resonance variable $\theta_3= 5\lambda - 3\lambda' - 2\varpi$ librates around $180^\circ$, see the right panel of Figure \ref{fig:2005TNNeptune}.

\subsubsection{Neptune - 136108~Haumea 12:7 MMR}

\begin{figure}
\includegraphics[width=\columnwidth]{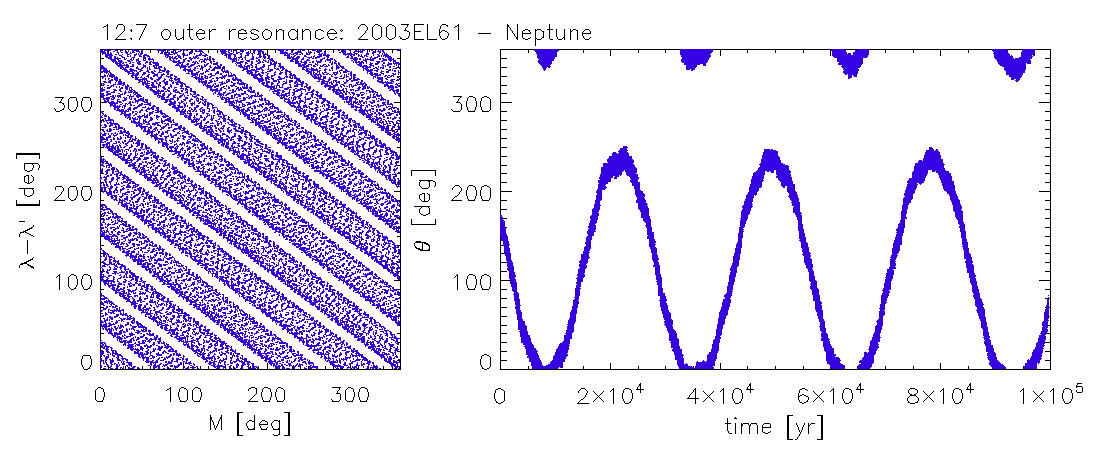}
\caption{Dwarf planet 136108~Haumea in a 12:7 MMR with Neptune}
\label{fig:haumeaneptune}
\end{figure}

136108~Haumea is a dwarf planet orbiting beyond Neptune, being discovered independently by \cite{2005ApJ...632L..45B} and \cite{2005MPEC....O...36O}. It is known that there is a 12:7 MMR of outer type between Haumea and Neptune. This is a 5th order MMR, and its identification might not be too straightforward being the ratio of the mean motions close to the 2:1 ratio. Seeking a resonant behaviour with Neptune of outer type, we plot $\lambda - \lambda'$ against $M$, see Figure \ref{fig:haumeaneptune}. As usual, $\lambda'$ refers to Neptune's mean orbital longitude, while the non-primed quantities to Haumea's orbital elements. Investigating Figure \ref{fig:haumeaneptune}, one can identify $q=5$ crossings with the horizontal, and $p=7$ crossings with the vertical axis. This yields to a mean motion ratio of $(p+q)/p = 12/7$, and a resonance variable $\theta_3 = 12\lambda - 7\lambda' - 5\varpi$ librating around approximately $120^\circ$. 

\subsubsection{Neptune - 2001~QR$_\mathrm{322}$ 1:1 MMR}

\begin{figure}
\includegraphics[width=\columnwidth]{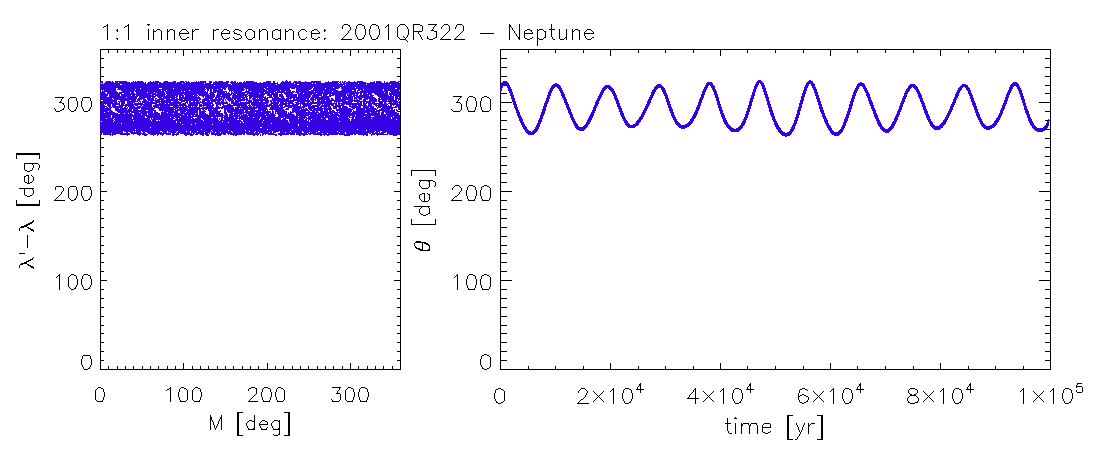}
\caption{Neptune and 2001QR$_\mathrm{322}$ in 1:1 MMR}
\label{fig:neptunetrojan}
\end{figure}

The method \emph{FAIR} can also be applied to Trojan-type motions. The main characteristic of the Trojan-type, or in other words co-orbital motion is the 1:1 MMR meaning that the bodies involved are sharing similar orbits. The best known examples for Trojan-type motion are the Trojan asteroids populating the neighbourhood of the stable triangular Lagrangian points $L_4$ and $L_5$ of the Sun-Jupiter system. The resonance variable in these cases is $\lambda-\lambda'$ which librates either around $60^\circ$, or $300^\circ$ depending on whether the asteroid is in the vicinity of the $L_4$, or $L_5$ point. In the $\lambda' - \lambda$ versus $M$ plots of the Neptune's Trojan 2001~QR (see Figure \ref{fig:neptunetrojan}, left panel) we find one strip parallel to the horizontal axis of the coordinate system yielding $q=0$ (see Table \ref{tab1}, first row), e.g. a zeroth order resonance. The only crossing with the vertical axis means $p+q=1$, that is $p=1$. Thus the resonance is $(p+q)/p=1/1$, and the resonance variable is $\theta = \lambda' - \lambda$, librating around $300^\circ$ (Figure \ref{fig:neptunetrojan}, right panel). 

\subsection{Mean motion resonances in exoplanetary systems}

Finally, with two additional examples we demonstrate the applicability of the method \emph{FAIR} to identify resonances between pairs of exoplanets, or chains of resonances in which many of them are involved. The numerical integration of orbits have been performed by using the already mentioned RKN 6/7 integrator.

\subsubsection{HD~60532}

\begin{figure}
\includegraphics[width=\columnwidth]{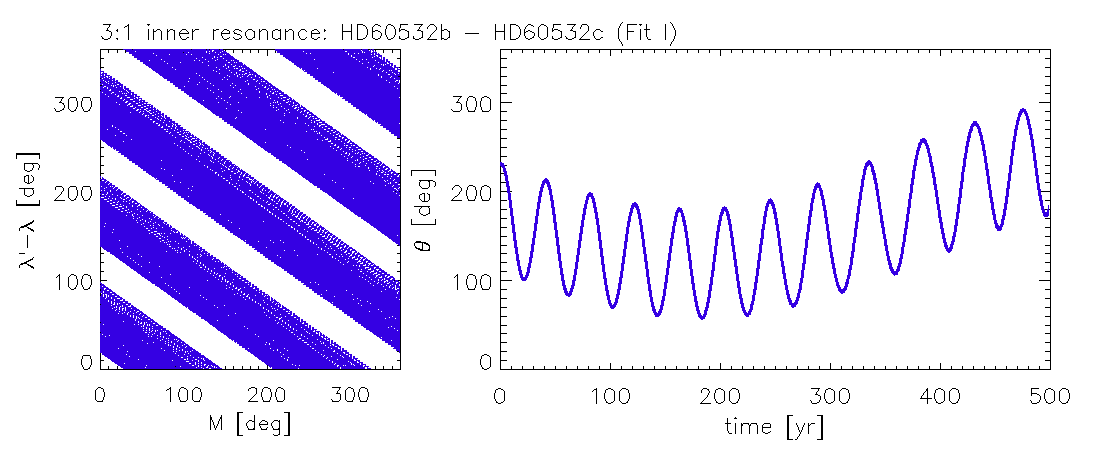}
\includegraphics[width=\columnwidth]{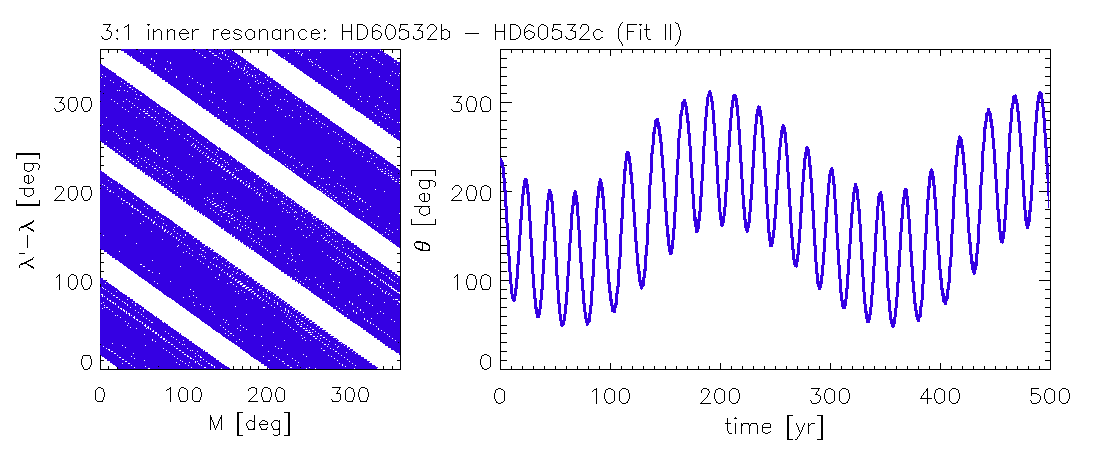}
\caption{The HD~60532 system in 3:1 inner MMR}
\label{fig:hd60532}
\end{figure}

Until the discovery of the system around HD~60532, an F-type star with mass $M_*=1.44~M_\odot$ by \cite{2008A&A...491..883D}, there were not known any giant planets involved in the 3:1 MMR. The final confirmation that giant planets can be in the 3:1 MMR was given by \cite{2009A&A...496L...5L}, while a formation study favouring a migration based scenario was performed by \cite{2010A&A...517A..31S}. In the latter study the authors demonstrated that the convergent migration, thus capture in the 3:1 MMR was only possible for larger planetary masses corresponding to Fit II of \cite{2009A&A...496L...5L}. In this work we checked both Fit~I with low and Fit~II with high planetary masses to show the efficiency of the method \emph{FAIR}. Our results are displayed in Figure \ref{fig:hd60532}, where in the upper left panel the $(\lambda'-\lambda)$ versus $M$ plot is displayed for Fit I, while in the bottom left panel for Fit II. These resonances are of inner types (as the primed orbital elements are refering to the outer giant planet), and by calculating the number of the crossings of the strips with the horizontal and vertical axis we have (see Table \ref{tab1}, first row) $q=2$ and $p+q=3$. Thus $p=1$, and the ratio of the mean motions is $(p+q)/p=3/1$. This means that without any a priori knowledge, just by a careful analysis of the $(\lambda'-\lambda)$ versus $M$ plot we could identify the mean motion resonance, and also write down the corresponding resonant variable, which in these cases is $\theta_1 = 3\lambda' - \lambda -2\varpi$, librating around $180^{\circ}$ (Figure \ref{fig:hd60532}, right panels). It is noteworthy that due to the relatively large libration amplitude, the strips in the $(\lambda'-\lambda)$ versus $M$ plot are quite broad. 

\subsubsection{Kepler~60}

\begin{figure}
\includegraphics[width=\columnwidth]{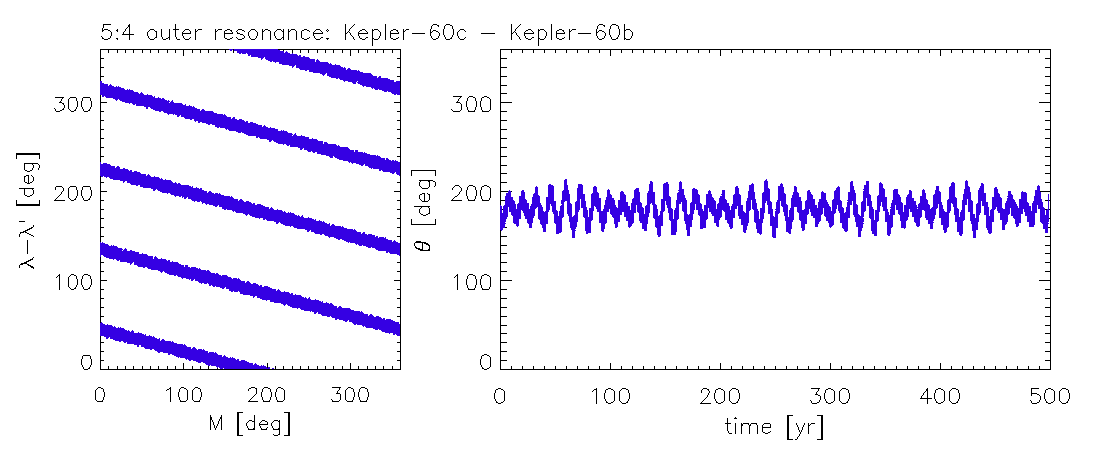}
\includegraphics[width=\columnwidth]{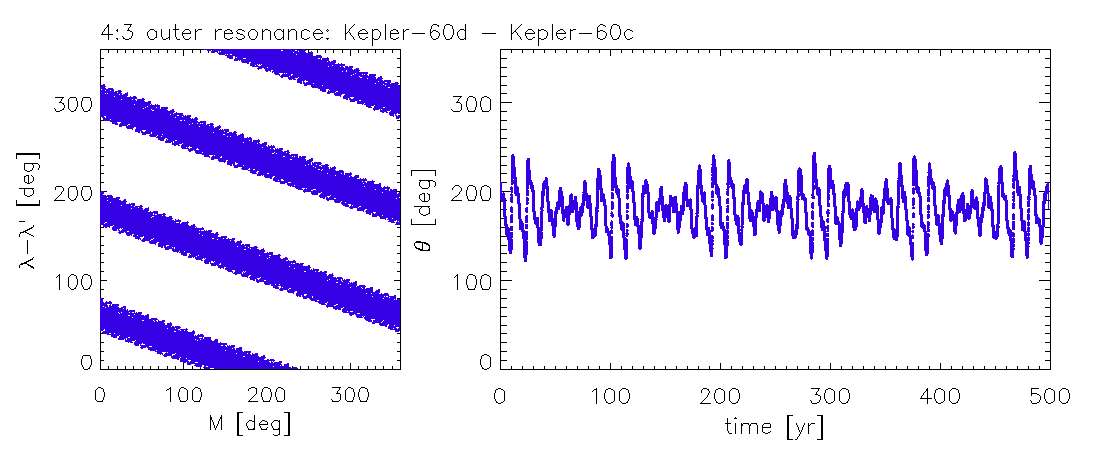}
\includegraphics[width=\columnwidth]{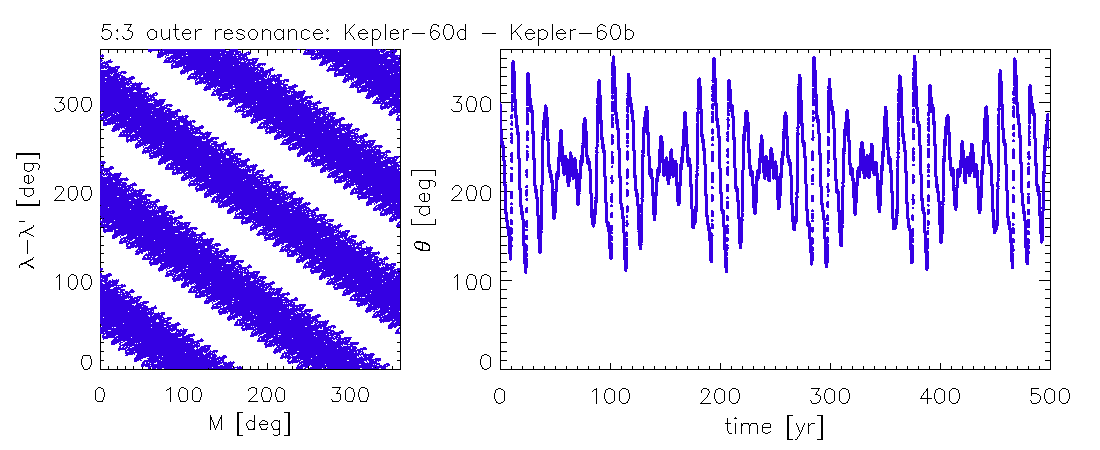}
\caption{The resonant pairs of the system Kepler 60, the three planets being
  in 5:4:3 Laplace-type MMR}
\label{fig:kep60}
\end{figure}

As we have already mentioned, there are planetary systems, typically discovered by the Kepler mission, in which the planets are captured in chains of mean motion resonances. A prominent example of these systems is Kepler~60, where three planets with masses around $4~M_\oplus$ are in the 5:4:3 Laplace-type MMR \citep{2016MNRAS.455L.104G}. Here we apply the method \emph{FAIR} to the Fit II of the cited paper, in which the pairs of the planets in MMRs are investigated. First, without any a priori knowlegde on the MMRs, we can study planets b and c. We display $\lambda_1 - \lambda_2$ as the function of $M_1$, see the upper left panel of Figure \ref{fig:kep60}. We note that the index 1 stays for the inner, while index 2 for the outer planet, planets b and c in this configuration. Considering that $\lambda$, $\varpi$, $M$ correspond to $\lambda_1$, $\varpi_1$, $M_1$ and $\lambda'$, $\varpi'$, $M'$ correspond to $\lambda_2$, $\varpi_2$, $M_2$, and that this is an outer type MMR, according to Table \ref{tab1} (third row) the numbers of crossings with the horizontal and vertical axis give $q=1$ and $p=4$. The mean motions ratio in this case is thus $(p+q)/p = 5/4$ corresponding to a resonant variable $\theta_{3,12} = 5\lambda_1 - 4\lambda_2-\varpi_1$, librating around $180^{\circ}$ (see the upper right panel of Figure \ref{fig:kep60}). 

As a next step, we can also check, whether the planets c and d are in a MMR too. To do so, we plot $\lambda_2 - \lambda_3$ as the function of $M_2$, shown in the middle left panel of Figure \ref{fig:kep60}. Now the non-primed elements correspond to those with index 2, and the primed ones to those with index 3. According to Table \ref{tab1} (third row), we have $q=1$ and $p=3$, that is there is a $(p+q)/p = 4/3$ MMR with a resonant variable $\theta_{3,23} = 4\lambda_2 - 3\lambda_3-\varpi_2$, librating around $180^{\circ}$ (middle right panel of Figure \ref{fig:kep60}).

Finally, we also plot $\lambda_1 - \lambda_3$ as the function of $M_1$ (bottom left panel of Figure \ref{fig:kep60}). Counting the numbers of crossings of the stripes with the axes, we find $q=2$ and $p=3$ corresponding to a mean motion ratio $(p+q)/p = 5/3$, and indeed, the resonant variable $\theta_{3,13}=5\lambda_1-3\lambda_3-2\varpi_1$ librates around $~225^\circ$
(bottom right panel of Figure \ref{fig:kep60}). 
Thus using the system Kepler~60, we gave an evidence that the method \emph{FAIR} is also applicable for a quick identification of MMRs between pairs of exoplanets being in resonant chains. The above analysis can also be performed for \emph{inner} type resonances. 

\subsection{Application of the method \emph{FAIR} to inclination-type MMR}

\begin{figure}
\includegraphics[width=\columnwidth]{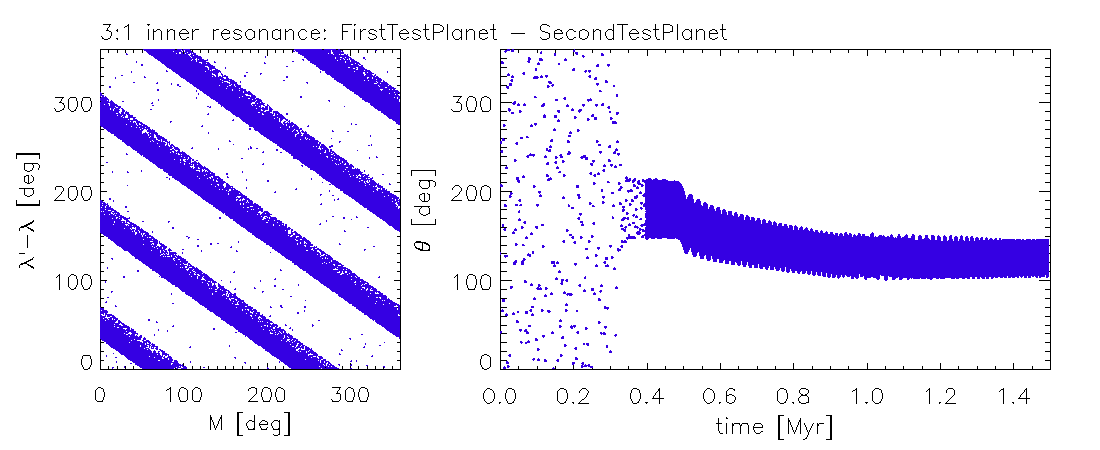}
\caption{Two migrating giant planets captured into a 3:1 eccentricity-type MMR. The method \emph{FAIR} is applied to the whole length of numerical integration.}
\label{fig:migr_giants31ecc_mmr}
\end{figure}

\begin{figure}
\includegraphics[width=\columnwidth]{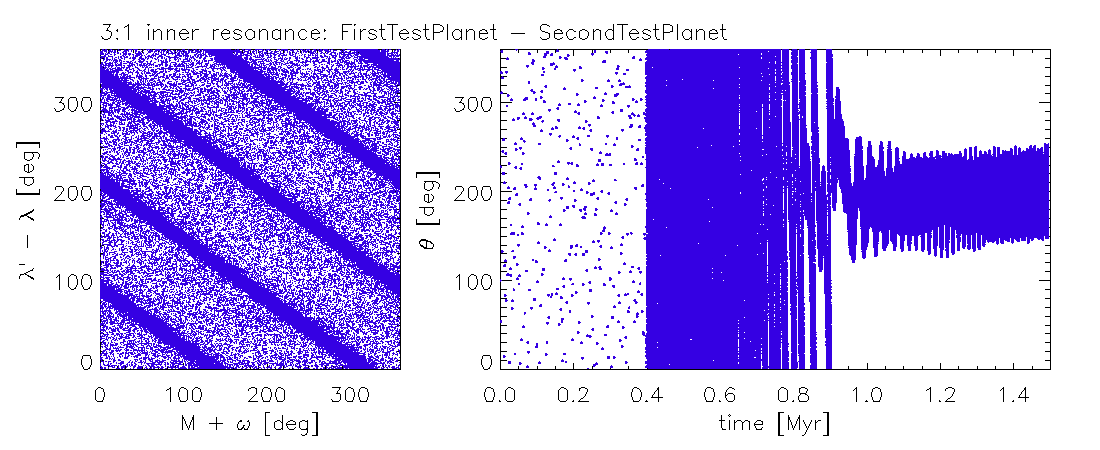}
\caption{Two migrating giant planets captured into a 3:1 inclination-type MMR. The method \emph{FAIR} is applied to the whole lenght of numerical integration.}
\label{fig:migr_giants31inc_mmr}
\end{figure}

Similarly to the eccentricity-type MMRs, the method \emph{FAIR} can also be applied to detect inclination-type resonances. The best known example for inclination-type MMR in the Solar System is between the Saturnian satellites Mimas and Thetis corresponding to a 2:1 commensurability. Inclination-type MMRs can also be developed between migrating giant planets if the eccentricity damping of the outer migrating planet is modest, allowing the rapid growth of the inner planet's eccentricity. This can additionally lead to the fast growth of the inner planet's inclination \citep[see][]{2009MNRAS.400.1373L}. Here we consider the following example that leads to a 3:1 MMR: the inner planet with mass $m_1=1M_J$ (where $M_J$ is the mass of Jupiter) is started from 5~au from nearly circular and planar, initially non-migrating orbit. An outer planet with mass $m_2=2M_J$ is started from 16.5~au from a circular orbit with negligible inclination with respect to the plane of reference of the coordinate system, and is forced to migrate in timescale of $\tau_{a}=7\times 10^5$ years, with the same eccentricity damping timescale. According to \cite{2009MNRAS.400.1373L}, the two planets enter first into a 3:1 eccentricity-type, and later on into an inclination-type MMR, as the libration of the following eccentricity-type and inclination-type variables clearly indicates:
\begin{equation}
\theta_1 = 3\lambda_2 - \lambda_1 -2\varpi_1,
\end{equation}
\begin{equation}
\theta_{I,1} = 3\lambda_2 - \lambda_1 -2\Omega_1.
\end{equation}
The resonance variable $\theta_1$ begins its libration roughly after $t \sim 3.5\times10^5$ years being the time when the resonant capture happens (see Figure \ref{fig:migr_giants31ecc_mmr}). Interestingly, the center of libration is shifted from $180^{\circ}$ to lower values between $100^{\circ} - 150^{\circ}$. The inclination of the inner planet gets excited around $t\sim 9.5\times 10^5$ years, that coincides to the libration of the resonance variable $\theta_{I,1}$ (see Figure \ref{fig:migr_giants31inc_mmr}). After this epoch the system is both in  3:1 eccentricity-type and inclination-type MMR, see the right panels of Figures \ref{fig:migr_giants31ecc_mmr} and \ref{fig:migr_giants31inc_mmr}. Studying the left panels of these figures that display the plots $(\lambda_2-\lambda_1)$ versus $M_1$ for the eccentricity-type, and $(\lambda_2-\lambda_1)$ versus $(M_1+\omega_1)$ for inclination-type MMRs, one can see that the method \emph{FAIR} is able to identify both eccentricity-type and inclination-type MMRs providing the correct resonance variables. We note that the plots have been made for the whole timespan of the numerical integration, also before the planets are got captured in resonances, thus there are scattered points in the figures being not yet settled in the stripes indicating the MMRs.

\subsection{Temporary capture into a MMR}
\begin{figure}
\includegraphics[width=\columnwidth]{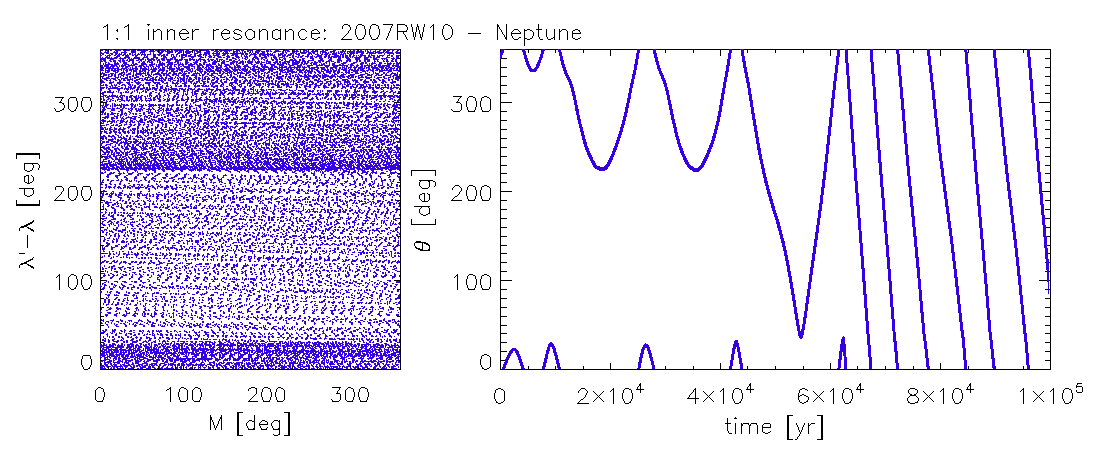}
\caption{Temporary capture of 2007~RW$_{10}$ into a 1:1 MMR, being a co-orbital companion of Neptune for a while.}
\label{fig:2007RW10_coorb_Neptune}
\end{figure}
Finally, we also show that the method \emph{FAIR} is able to detect temporary capture into a MMR. It can happen, for instance, that during their migration planets enter into a MMR, see our results in the previous part. The left panels of Figures \ref{fig:migr_giants31ecc_mmr} and \ref{fig:migr_giants31inc_mmr} contain scattered points corresponding to that epochs when the planets are not engaged into the 3:1 MMR. On the other hand it is important that the scattered points should not repress those settled in the stripes. 

Beside the cases of migrating planets, there are examples of temporary capture of bodies in co-orbital orbits, too. This happens to the asteroid 2007~RW$_{10}$, which was assumed as Neptune Trojan \citep{2012A&A...545L...9D}. The right panel of Figure \ref{fig:2007RW10_coorb_Neptune} shows the temporary libration of the synodic longitude $\lambda'-\lambda$, around $300^\circ$. The left panel of this figure shows the $(\lambda'-\lambda)$ versus $M$ plot, in which one parallel stripe to the horizontal axis becomes visible around $\lambda'-\lambda= 300^\circ$ among the scattered points indicating the case of the 1:1 MMR, see additionally Figure \ref{fig:neptunetrojan}.

\section{Conclusions}
In this paper we presented a novel method which is suitable to easily decide whether two planets are involved in some mean motion resonance, being either of eccentricity or inclination type, without any a priori knowledge of its character. Based on the geometrical meaning of the resonance variables, first we described in detail how our method works. Next we demonstrated through a few examples of real bodies of the Solar System and of exoplanetary systems, involved in various MMRs, the straightforward applicability and efficiency of the method \emph{FAIR}. We have found that this method is able to detect MMRs also in those cases when the involved bodies are only temporarily captured, such as migrating pairs of planets or co-orbital companions to giant planets. Our method can also serve as a practical help for future analysis of celestial bodies which may be envolved in mean motion resonances.

\section*{Acknowledgements}

We thank the anonymous referee for her/his suggestions and comments helping us to improve the work. This research has been supported by the Hungarian National Research, Development and Innovation Office, NKFIH grant K-119993 and the HAS Wigner RCP -- GPU-Lab.
Zs. S\'andor thanks the support of the J\'anos Bolyai Research Scholarship of the Hungarian Academy of Sciences.

%%%%%%%%%%%%%%%%%%%%%%%%%%%%%%%%%%%%%%%%%%%%%%%%%%



%\bibliographystyle{mnras}

\bibliography{Forgacs-Dajka-et-al-FAIR}

\begin{thebibliography}{}
\makeatletter
\relax
\def\mn@urlcharsother{\let\do\@makeother \do\$\do\&\do\#\do\^\do\_\do\%\do\~}
\def\mn@doi{\begingroup\mn@urlcharsother \@ifnextchar [ {\mn@doi@}
  {\mn@doi@[]}}
\def\mn@doi@[#1]#2{\def\@tempa{#1}\ifx\@tempa\@empty \href
  {http://dx.doi.org/#2} {doi:#2}\else \href {http://dx.doi.org/#2} {#1}\fi
  \endgroup}
\def\mn@eprint#1#2{\mn@eprint@#1:#2::\@nil}
\def\mn@eprint@arXiv#1{\href {http://arxiv.org/abs/#1} {{\tt arXiv:#1}}}
\def\mn@eprint@dblp#1{\href {http://dblp.uni-trier.de/rec/bibtex/#1.xml}
  {dblp:#1}}
\def\mn@eprint@#1:#2:#3:#4\@nil{\def\@tempa {#1}\def\@tempb {#2}\def\@tempc
  {#3}\ifx \@tempc \@empty \let \@tempc \@tempb \let \@tempb \@tempa \fi \ifx
  \@tempb \@empty \def\@tempb {arXiv}\fi \@ifundefined
  {mn@eprint@\@tempb}{\@tempb:\@tempc}{\expandafter \expandafter \csname
  mn@eprint@\@tempb\endcsname \expandafter{\@tempc}}}

\bibitem[\protect\citeauthoryear{{Bro{\v z}} \& {Vokrouhlick{\'y}}}{{Bro{\v z}}
  \& {Vokrouhlick{\'y}}}{2008}]{2008MNRAS.390..715B}
{Bro{\v z}} M.,  {Vokrouhlick{\'y}} D.,  2008, \mn@doi [\mnras]
  {10.1111/j.1365-2966.2008.13764.x}, \href
  {http://adsabs.harvard.edu/abs/2008MNRAS.390..715B} {390, 715}

\bibitem[\protect\citeauthoryear{{Brown} et~al.,}{{Brown}
  et~al.}{2005}]{2005ApJ...632L..45B}
{Brown} M.~E.,  et~al., 2005, \mn@doi [\apjl] {10.1086/497641}, \href
  {http://adsabs.harvard.edu/abs/2005ApJ...632L..45B} {632, L45}

\bibitem[\protect\citeauthoryear{{Cohen} \& {Hubbard}}{{Cohen} \&
  {Hubbard}}{1965}]{1965AJ.....70...10C}
{Cohen} C.~J.,  {Hubbard} E.~C.,  1965, \mn@doi [\aj] {10.1086/109674}, \href
  {http://adsabs.harvard.edu/abs/1965AJ.....70...10C} {70, 10}

\bibitem[\protect\citeauthoryear{{Desort}, {Lagrange}, {Galland}, {Beust},
  {Udry}, {Mayor}  \& {Lo Curto}}{{Desort} et~al.}{2008}]{2008A&A...491..883D}
{Desort} M.,  {Lagrange} A.-M.,  {Galland} F.,  {Beust} H.,  {Udry} S.,
  {Mayor} M.,   {Lo Curto} G.,  2008, \mn@doi [\aap]
  {10.1051/0004-6361:200810241}, \href
  {http://adsabs.harvard.edu/abs/2008A%26A...491..883D} {491, 883}

\bibitem[\protect\citeauthoryear{{Dormand} \& {Prince}}{{Dormand} \&
  {Prince}}{1978}]{1978CeMec..18..223D}
{Dormand} J.~R.,  {Prince} P.~J.,  1978, \mn@doi [Celestial Mechanics]
  {10.1007/BF01230162}, \href
  {http://adsabs.harvard.edu/abs/1978CeMec..18..223D} {18, 223}

\bibitem[\protect\citeauthoryear{{Galiazzo}, {Bazs{\'o}}  \&
  {Dvorak}}{{Galiazzo} et~al.}{2013}]{2013P&SS...84....5G}
{Galiazzo} M.~A.,  {Bazs{\'o}} {\'A}.,   {Dvorak} R.,  2013, \mn@doi [\planss]
  {10.1016/j.pss.2013.03.017}, \href
  {http://adsabs.harvard.edu/abs/2013P%26SS...84....5G} {84, 5}

\bibitem[\protect\citeauthoryear{{Go{\'z}dziewski} \&
  {Migaszewski}}{{Go{\'z}dziewski} \&
  {Migaszewski}}{2014}]{2014MNRAS.440.3140G}
{Go{\'z}dziewski} K.,  {Migaszewski} C.,  2014, \mn@doi [\mnras]
  {10.1093/mnras/stu455}, \href
  {http://adsabs.harvard.edu/abs/2014MNRAS.440.3140G} {440, 3140}

\bibitem[\protect\citeauthoryear{{Go{\'z}dziewski}, {Migaszewski}, {Panichi}
  \& {Szuszkiewicz}}{{Go{\'z}dziewski} et~al.}{2016}]{2016MNRAS.455L.104G}
{Go{\'z}dziewski} K.,  {Migaszewski} C.,  {Panichi} F.,   {Szuszkiewicz} E.,
  2016, \mn@doi [\mnras] {10.1093/mnrasl/slv156}, \href
  {http://adsabs.harvard.edu/abs/2016MNRAS.455L.104G} {455, L104}

\bibitem[\protect\citeauthoryear{{Hahn} \& {Malhotra}}{{Hahn} \&
  {Malhotra}}{2005}]{2005AJ....130.2392H}
{Hahn} J.~M.,  {Malhotra} R.,  2005, \mn@doi [\aj] {10.1086/452638}, \href
  {http://adsabs.harvard.edu/abs/2005AJ....130.2392H} {130, 2392}

\bibitem[\protect\citeauthoryear{{Laskar} \& {Correia}}{{Laskar} \&
  {Correia}}{2009}]{2009A&A...496L...5L}
{Laskar} J.,  {Correia} A.~C.~M.,  2009, \mn@doi [\aap]
  {10.1051/0004-6361/200911689}, \href
  {http://adsabs.harvard.edu/abs/2009A%26A...496L...5L} {496, L5}

\bibitem[\protect\citeauthoryear{{Levison}, {Morbidelli}, {Van Laerhoven},
  {Gomes}  \& {Tsiganis}}{{Levison} et~al.}{2008}]{2008Icar..196..258L}
{Levison} H.~F.,  {Morbidelli} A.,  {Van Laerhoven} C.,  {Gomes} R.,
  {Tsiganis} K.,  2008, \mn@doi [\icarus] {10.1016/j.icarus.2007.11.035}, \href
  {http://adsabs.harvard.edu/abs/2008Icar..196..258L} {196, 258}

\bibitem[\protect\citeauthoryear{{Libert} \& {Tsiganis}}{{Libert} \&
  {Tsiganis}}{2009}]{2009MNRAS.400.1373L}
{Libert} A.-S.,  {Tsiganis} K.,  2009, \mn@doi [\mnras]
  {10.1111/j.1365-2966.2009.15532.x}, \href
  {http://adsabs.harvard.edu/abs/2009MNRAS.400.1373L} {400, 1373}

\bibitem[\protect\citeauthoryear{{Marcy}, {Butler}, {Fischer}, {Vogt},
  {Lissauer}  \& {Rivera}}{{Marcy} et~al.}{2001}]{2001ApJ...556..296M}
{Marcy} G.~W.,  {Butler} R.~P.,  {Fischer} D.,  {Vogt} S.~S.,  {Lissauer}
  J.~J.,   {Rivera} E.~J.,  2001, \mn@doi [\apj] {10.1086/321552}, \href
  {http://adsabs.harvard.edu/abs/2001ApJ...556..296M} {556, 296}

\bibitem[\protect\citeauthoryear{{Mills}, {Fabrycky}, {Migaszewski}, {Ford},
  {Petigura}  \& {Isaacson}}{{Mills} et~al.}{2016}]{2016Natur.533..509M}
{Mills} S.~M.,  {Fabrycky} D.~C.,  {Migaszewski} C.,  {Ford} E.~B.,  {Petigura}
  E.,   {Isaacson} H.,  2016, \mn@doi [\nat] {10.1038/nature17445}, \href
  {http://adsabs.harvard.edu/abs/2016Natur.533..509M} {533, 509}

\bibitem[\protect\citeauthoryear{{Murray} \& {Dermott}}{{Murray} \&
  {Dermott}}{1999}]{1999ssd..book.....M}
{Murray} C.~D.,  {Dermott} S.~F.,  1999, {Solar system dynamics}

\bibitem[\protect\citeauthoryear{{Ortiz} et~al.,}{{Ortiz}
  et~al.}{2005}]{2005MPEC....O...36O}
{Ortiz} J.~L.,  et~al., 2005, Minor Planet Electronic Circulars, \href
  {http://adsabs.harvard.edu/abs/2005MPEC....O...36O} {2005-O36}

\bibitem[\protect\citeauthoryear{{Rivera}, {Laughlin}, {Butler}, {Vogt},
  {Haghighipour}  \& {Meschiari}}{{Rivera} et~al.}{2010}]{2010ApJ...719..890R}
{Rivera} E.~J.,  {Laughlin} G.,  {Butler} R.~P.,  {Vogt} S.~S.,  {Haghighipour}
  N.,   {Meschiari} S.,  2010, \mn@doi [\apj] {10.1088/0004-637X/719/1/890},
  \href {http://adsabs.harvard.edu/abs/2010ApJ...719..890R} {719, 890}

\bibitem[\protect\citeauthoryear{{S{\'a}ndor} \& {Kley}}{{S{\'a}ndor} \&
  {Kley}}{2010}]{2010A&A...517A..31S}
{S{\'a}ndor} Z.,  {Kley} W.,  2010, \mn@doi [\aap]
  {10.1051/0004-6361/201014072}, \href
  {http://adsabs.harvard.edu/abs/2010A%26A...517A..31S} {517, A31}

\bibitem[\protect\citeauthoryear{{Sheppard}, {Trujillo}  \&
  {Marsden}}{{Sheppard} et~al.}{2005}]{2005MPEC....U...97S}
{Sheppard} S.~S.,  {Trujillo} C.~A.,   {Marsden} B.~G.,  2005, Minor Planet
  Electronic Circulars, \href
  {http://adsabs.harvard.edu/abs/2005MPEC....U...97S} {2005-U97}

\bibitem[\protect\citeauthoryear{{Tinney}, {Butler}, {Marcy}, {Jones},
  {Laughlin}, {Carter}, {Bailey}  \& {O'Toole}}{{Tinney}
  et~al.}{2006}]{2006ApJ...647..594T}
{Tinney} C.~G.,  {Butler} R.~P.,  {Marcy} G.~W.,  {Jones} H.~R.~A.,  {Laughlin}
  G.,  {Carter} B.~D.,  {Bailey} J.~A.,   {O'Toole} S.,  2006, \mn@doi [\apj]
  {10.1086/503706}, \href {http://adsabs.harvard.edu/abs/2006ApJ...647..594T}
  {647, 594}

\bibitem[\protect\citeauthoryear{{Vogt}, {Butler}, {Marcy}, {Fischer}, {Henry},
  {Laughlin}, {Wright}  \& {Johnson}}{{Vogt}
  et~al.}{2005}]{2005ApJ...632..638V}
{Vogt} S.~S.,  {Butler} R.~P.,  {Marcy} G.~W.,  {Fischer} D.~A.,  {Henry}
  G.~W.,  {Laughlin} G.,  {Wright} J.~T.,   {Johnson} J.~A.,  2005, \mn@doi
  [\apj] {10.1086/432901}, \href
  {http://adsabs.harvard.edu/abs/2005ApJ...632..638V} {632, 638}

\bibitem[\protect\citeauthoryear{{de la Fuente Marcos} \& {de la Fuente
  Marcos}}{{de la Fuente Marcos} \& {de la Fuente
  Marcos}}{2012}]{2012A&A...545L...9D}
{de la Fuente Marcos} C.,  {de la Fuente Marcos} R.,  2012, \mn@doi [\aap]
  {10.1051/0004-6361/201219931}, \href
  {http://adsabs.harvard.edu/abs/2012A%26A...545L...9D} {545, L9}

\makeatother
\end{thebibliography}

%\begin{thebibliography}{99}
%\end{thebibliography}

%%%%%%%%%%%%%%%%%%%%%%%%%%%%%%%%%%%%%%%%%%%%%%%%%%

%%%%%%%%%%%%%%%%% APPENDICES %%%%%%%%%%%%%%%%%%%%%

%\appendix



%%%%%%%%%%%%%%%%%%%%%%%%%%%%%%%%%%%%%%%%%%%%%%%%%%

% Don't change these lines
\bsp	% typesetting comment
\label{lastpage}
\end{document}